# Investigation of Electrical Conductivity Changes during Brain Functional Activity in 3T MRI

Kyu-Jin Jung[1], Chuanjiang Cui[1], Soo-Hyung Lee[1], Chan-Hee Park[1], Ji-Won Chun[2] and Dong-Hyun Kim[1]

[1]*Department of Electrical and Electronic Engineering, Yonsei University, Seoul, Republic of Korea*

[2]*Department of Medical Informatics, Catholic University of Korea College of Medicine, Seoul, Republic of Korea*

**Correspondence to:**

Dong-Hyun Kim, Ph.D.

Department of Electrical and Electronic Engineering

Yonsei University

Seoul, Korea

E-mail: *donghyunkim@yonsei.ac.kr*

# Abstract

Blood oxygenation level-dependent (BOLD) functional magnetic resonance imaging (fMRI) is widely used to visualize brain activation regions by detecting hemodynamic responses associated with increased metabolic demand. While alternative MRI methods have been employed to monitor functional activities, the investigation of in-vivo electrical property changes during brain function remains limited. In this study, we explored the relationship between fMRI signals and electrical conductivity (measured at the Larmor frequency) changes using phase-based electrical properties tomography (EPT).

Our results revealed consistent patterns: conductivity changes showed negative correlations, with conductivity decreasing in the functionally active regions whereas $B_1$ phase mapping exhibited positive correlations around activation regions. These observations were consistent across both motor and visual cortex activations. To further substantiate these findings, we conducted electromagnetic radio-frequency simulations that modeled activation states with varying conductivity, which demonstrated trends similar to our in-vivo results for both $B_1$ phase and conductivity.

These findings suggest that in-vivo electrical conductivity changes can indeed be measured during brain activity. However, further investigation is needed to fully understand the underlying mechanisms driving these measurements.

# 1. Introduction

Non-invasive blood oxygenation level-dependent (BOLD) functional magnetic resonance imaging (fMRI) offers insights into brain activity[1,2]. It is known that the MR imaging source of the BOLD effect is attributed to changes in local magnetic fields (derived from susceptibility variation) caused by variations in the levels of oxygenated and deoxygenated hemoglobin in the blood tissue. The magnetic field can be detected through various MRI methods: gradient echo-planar imaging (GE-EPI) sequence[1,2], spin-echo echo-planar imaging (SE-EPI) sequence[3,4], and balanced steady-state free precession (bSSFP) sequence[5,6,7], to name a few. While MRI images contain both magnitude and phase components, the fMRI modality primarily relies on the MRI magnitude information to extract activation signals.

Particularly, GE-EPI is well-known for its sensitivity to the BOLD effect and is widely used[2,8,9]. Considerable efforts have been dedicated to elucidating the physiological mechanisms underlying the MR imaging source of the BOLD effect[10,11,12]. In addition, research[13,14] has been conducted in identifying quantitative changes in magnetic susceptibility by observing MRI phase changes. These approaches[13,14] leverage the principle that quantitative biophysical information can be derived from $B_0$ phase information, which encapsulates changes induced by the magnetic properties of in-vivo tissues. Functional quantitative susceptibility mapping[15,16,17] (QSM), based on $B_0$ phase information related to $T_2$* decay, enhances the detection of local magnetic susceptibility shifts and allows quantitative tracking of these susceptibility changes.

On the other hand, neuronal activity can impact the fMRI signals observed in sequences such as SE-EPI[3,4] and bSSFP[5,6]. These sequences mitigate dephasing effects and spatial distortion, potentially altering the sensitivity and representation of fMRI contrasts. Notably, susceptibility changes based on $T_2$* decay and $B_0$ phase cannot fully explain the MR imaging source of fMRI observed in bSSFP and SE-EPI sequences, which are predominantly influenced by $T_2$ decay and radio-frequency (RF) phase information ($B_1$ phase). To identify the characteristics of fMRI signal differences according to MRI modalities, the sensitivity to physiological variables, including vessel size, orientation, blood volume fraction, and magnetic field strength, has been investigated across GE-EPI, SE-EPI, and bSSFP sequences[7,18]. The results demonstrated that GE-EPI is more sensitive to larger vessels due to its susceptibility to intravascular effects, while SE-EPI showed a stronger dependence on smaller vessel contributions, particularly in regions with high blood volume fractions. In contrast, bSSFP exhibited a unique sensitivity to both intra/extra-vascular components, with its signal influenced by vessel orientation relative to the main magnetic field.

While changes in magnetic properties associated with activation have been relatively well studied using the aforementioned methods, alterations in electrical properties remain underexplored. Neuronal firing directly induces flux changes of ions in the intra- and extra-cellular domains, leading to modifications in electrical properties. However, in-vivo observations of these changes have not yet been demonstrated. A deeper understanding of electrical property changes could enhance the interpretation of fMRI signals, offering a more comprehensive view of neural activity by integrating both hemodynamic and electro-physiological information. Given this potential, several recent studies[19,20,21,22,23,24,25] have proposed the potential existence of functional activity related signal changes to electrical conductivity. Attempts have been made to investigate functional activity related changes associated with $B_1$ phase based on Electrical Properties Tomography[26,27] (EPT) algorithm. The $B_1$ phase is typically used for conductivity reconstruction known as phase-based EPT[27]. However, these findings have so far presented conflicting results regarding the relationship between fMRI signals and electrical conductivity changes. Specifically, as activation is triggered, both increases and decreases in conductivity have been reported, leading to ambiguity in interpretation.

Given the lack of in-vivo investigation of electrical property changes during function and the ambiguity among these studies, the purpose of this study is to determine the changes in reconstructed conductivity with functional activation. These changes were observed in response to stimuli (finger tapping and visual stimulation). We were able to observe significant fMRI-correlated signals in the $B_1$ phase and conductivity, while mitigating the potential for detecting spurious false-positive activation signals. Furthermore, we employed finite-difference time-domain (FDTD) RF simulations to validate the in-vivo observations of the activation signal changes in $B_1$ phase and conductivity.

# Results

## In-vivo Motor Cortex Stimulation Response using fMRI, $B_1$ phase, and Conductivity

The purpose of the in-vivo experiment 1 was to determine whether changes in $B_1$ phase occurred due to stimulation and to subsequently monitor the corresponding changes in conductivity on the activated regions. GE-EPI scan was performed with the number of signal averaging (NSA) = 1 and bSSFP scan with NSA=24 during a right-hand finger-tapping task. For the phase-based EPT algorithm, we employed the weighted polynomial fitting[28] (Poly-Fit) method (kernel size = $17\times17$). Details of the experimental setting are described in the Methods section.

We first observed the fMRI activation maps according to GE-EPI BOLD (Fig. 1A) and bSSFP (Fig. 1B). For both imaging sequences, the fMRI effect was noticeable in the left motor cortex area due to the right-hand finger-tapping stimulus ($p$<0.03), while the prominent activation regions varied subtly depending on the MRI modality. We analyzed the mean with the standard deviation (STD) signals in both right and left motor cortex for all volunteers ($n$=10). Regardless of the imaging sequence, it was observed that the activation signal amplitude (calculated as the average signal difference between offset and onset of stimulation) was prominent in the left motor cortex; GE-EPI: $0.0074 \pm 0.0021$ [a.u.] versus bSSFP: $0.0018 \pm 0.0004$ [a.u.] ($p$<0.03). GE-EPI demonstrated increased fMRI signal change sensitivity compared to bSSFP; GE-EPI: $1.1757 \pm 0.2316$ [%] versus bSSFP: $0.4505 \pm 0.1091$ [%]. The difference in STD was attributed to the varying NSA levels between the two MRI modalities. On the other hand, in the right motor cortex region, no distinct signal was detected in response to the finger-tapping stimulus.

Secondly, we examined whether the right-hand finger-tapping stimulus led to localized changes in $B_1$ phase measurements around the left motor cortex using bSSFP ($n$=10). The activation maps from $B_1$ phase information are presented in Fig. 2A. All phase activation maps consistently exhibited the dominant positively correlated activation regions surrounding the activation region. However, the activation regions in the $B_1$ phase for all volunteers did not completely coincide with the BOLD and bSSFP fMRI activation regions. In two volunteers (volunteers 8 and 9), spurious activation was observable near the paracentral lobule (see Fig. 2A, orange arrow). Significant positive phase changes were observed in the left motor cortex, while there were no significant phase changes in the right motor cortex. The activation signal amplitude from $B_1$ phase (calculated as the average signal difference between offset and onset of stimulation from $B_1$ phase signals) was calculated in both right and left motor cortex for all volunteers (*n*=10); left motor cortex (activation): $0.00030 \pm 0.00015$ [radian] versus right motor cortex (control): $0.00007 \pm 0.00022$ [radian] (*r*<0.03). Compared to the bSSFP magnitude signals, the fMRI signal change of the $B_1$ phase was smaller due to its susceptibility to noise; left motor cortex (activation): $0.0547 \pm 0.0306$ [%] versus right motor cortex (control): $0.0134 \pm 0.0295$ [%] (*r*<0.03). Despite the noise contamination, the bSSFP scan demonstrated that the amplitude of the activation signal from $B_1$ phase can exceed the noise STD, showing a prominent increasing trend in value in response to stimulation.

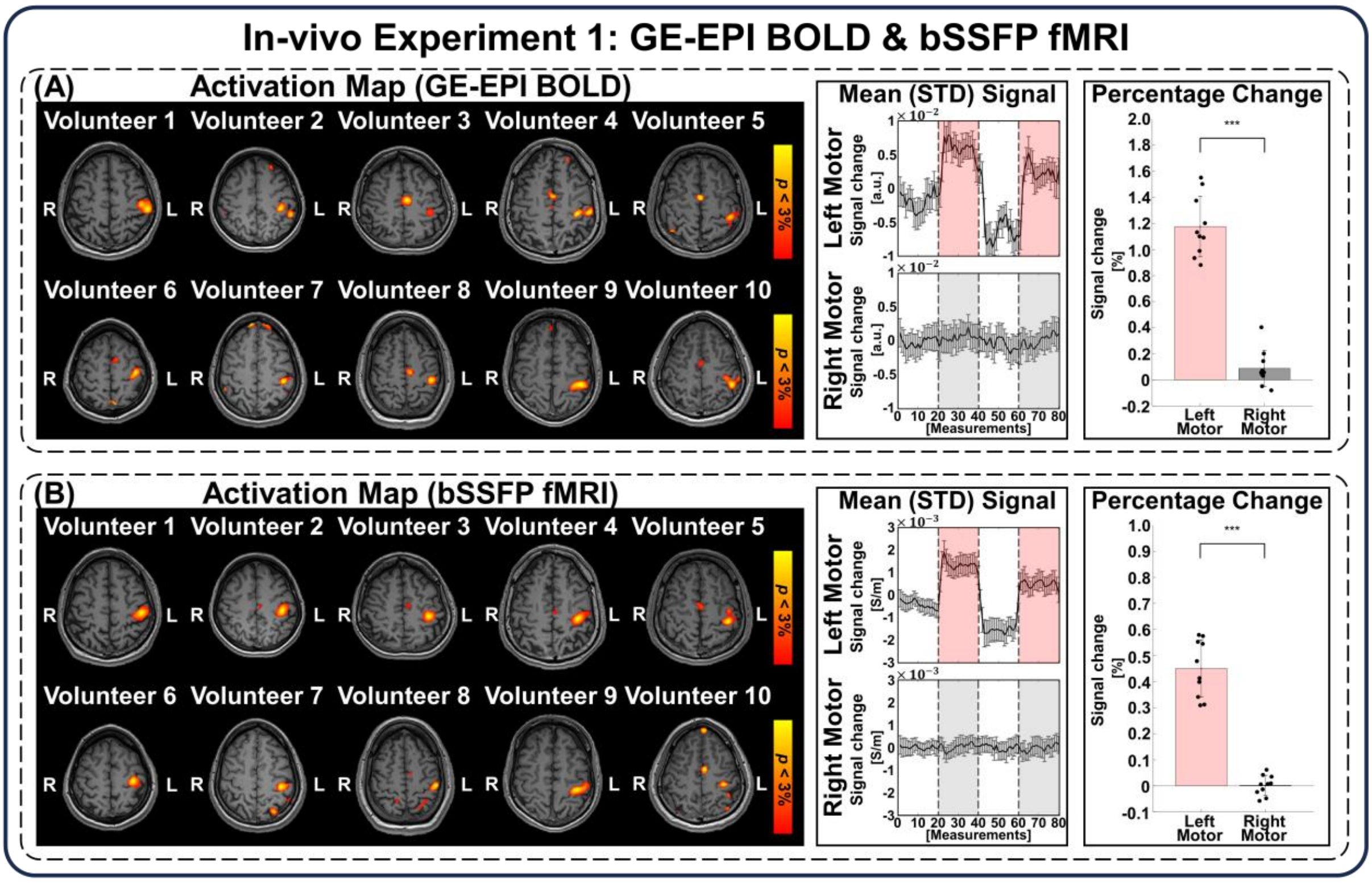

**Fig. 1|** In-vivo experiment 1: Activation maps of GE-EPI BOLD and bSSFP fMRI with mean (STD) signal and activation percentage during finger-tapping stimulus (*n*=10). **A** Activation maps of GE-EPI BOLD The BOLD signals were detected at a significance level of *p*<0.03. Mean (STD) temporal-series signal derived from the right (control) and left motor (activation) regions. For the percentage change analysis, *** indicates statistical significance of *p*<0.001. **B** Activation maps of the bSSFP fMRI. The bSSFP fMRI signals were identified at a significance level of *p*<0.03. Mean (STD) temporal-series signal derived from the right (control) and left motor cortex (activation) regions.

We next explored the conductivity changes around the motor cortex (Fig. 2B). Visual observation of the conductivity-based activation maps consistently indicated that conductivity changes were predominantly associated with negative correlation with respect to activation.

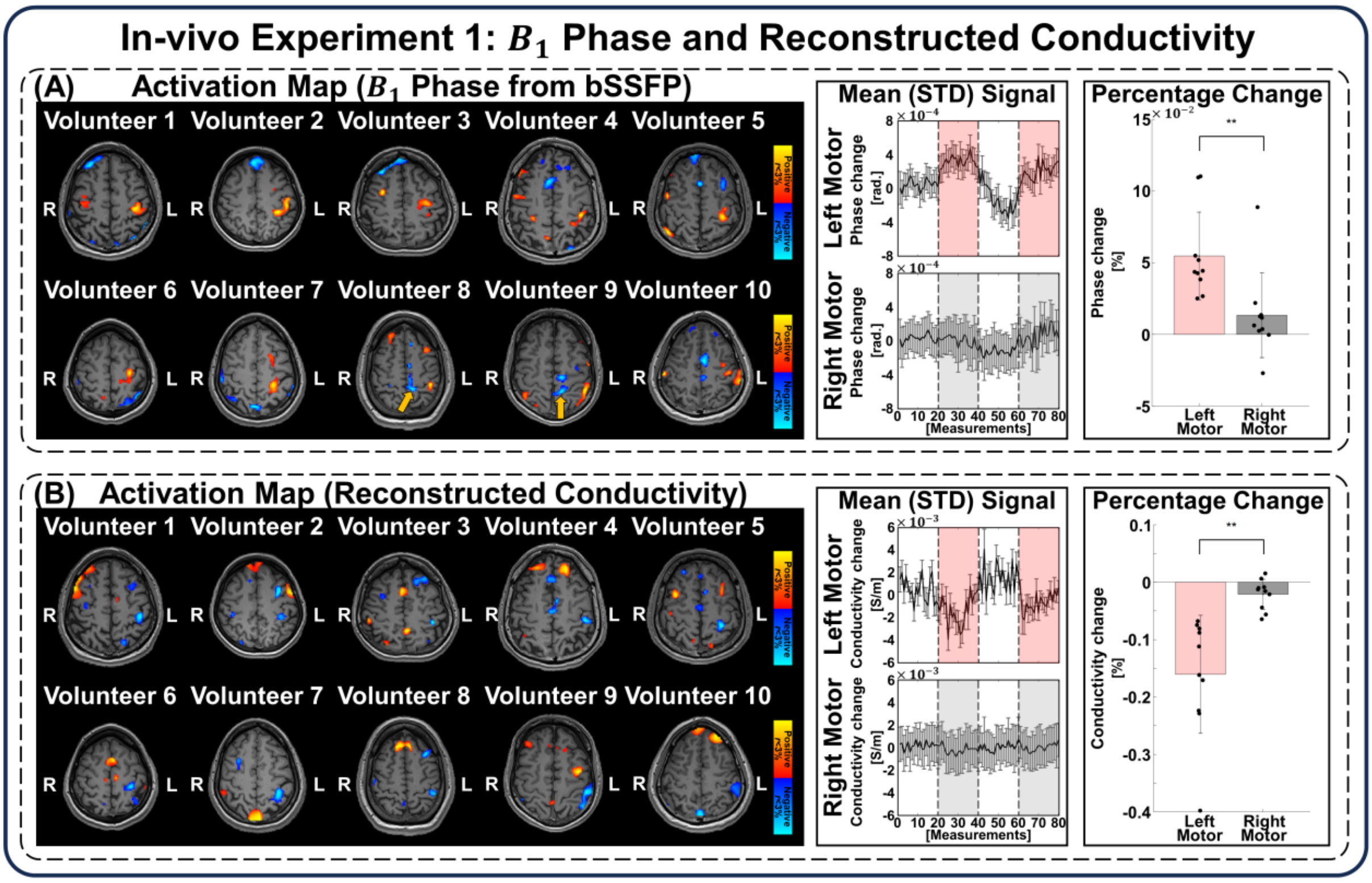

**Fig. 2|** In-vivo experiment 1: activation maps of $B_1$ phase and reconstructed conductivity with mean (STD) signal and activation percentage during finger-tapping stimulus in target regions ($n$=10). A correlation analysis strategy based on the block design was used to extract the activation map from $B_1$ phase and conductivity. Both positive and negative correlation coefficients were considered as significant, utilizing a two-tailed significance threshold of $r$<0.03. **A** Activation maps of the $B_1$ phase obtained from bSSFP. Mean (STD) temporal-series signal derived from the right (control) and left motor cortex (activation) regions. For the percentage change analysis, ** indicates statistical significance of $p$<0.01. **B** Activation maps of the reconstructed conductivity obtained from bSSFP phase. For the phase-based conductivity reconstruction, the Poly-Fit method was utilized. Mean (STD) temporal-series signals derived from the right (control) and left motor cortex (activation) regions.

Similar to the $B_1$ phase observations, the conductivity activation regions also did not completely coincide with the BOLD and bSSFP fMRI activation area. On the other hand, positively correlated activations were observed in inconsistent regions among volunteers. Nevertheless, in the left motor cortex region, significant negative conductivity changes were observed ($r$<0.03). The activation signal amplitude of reconstructed conductivity (calculated as the average signal difference between offset and onset of stimulation for conductivity values) was estimated in both the right and left motor cortex regions for all volunteers ($n$=10). For the activated left motor cortex, the values were $-0.0019 \pm 0.0010$ [S/m] and $-0.1605 \pm 0.1028$ [%], while for the right motor cortex (control), the values were $-0.0002 \pm 0.0016$ [S/m] and $-0.0211 \pm 0.0260$ [%], respectively. Similar to the $B_1$ phase observations, although the STD of the noise was amplified during the conductivity reconstruction process, the difference in conductivity due to activation remained higher than the noise STD, indicated by a decreasing trend in activation following the onset of the finger-tapping stimulation. Conversely, no significant changes were observed in the right motor cortex (though a comparison between the right and left motor cortices revealed a statistically significant difference, $p$<0.01, **). Additional results based on SE-EPI are summarized in Fig. S1.

## In-vivo Sensitivity of Motor Cortex Stimulation Response using $B_1$ phase and Conductivity Depending on the Number of Signal-Averaging

Given the sensitivity of $B_1$ phase and phase-based conductivity reconstruction algorithms to noise, and the potential for spurious false-positive activation, we performed in-vivo experiment 2, which examined the response to the right-hand finger-tapping stimulus under high signal-to-noise ratio (SNR) levels (Total NSA=192). The activation maps from the $B_1$ phase under various NSA levels are presented in Fig. 3A (NSA=12, 24, 48, 72, 144, 168, and 192; estimated SNR=354, 501, 756, 892, 1241, 1352, and 1487, respectively, computed based on the WM region). All activation maps from the $B_1$ phase across different NSA levels consistently exhibited the same positive correlation in the activation regions surrounding the left motor cortex. While spurious activation regions appeared near the left motor cortex (with both positively and negatively correlated $B_1$ phase changes) at NSA=12, negatively correlated activation ceased to be observed around the motor cortex region from NSA=24 and above. Additionally, in the extracted activation region of the $B_1$ phase, immediate responses to the stimulation were observed at NSA levels near 72 and above, compared to slower responses shown in Fig. 2. As the NSA level decreased, signal fluctuations became more pronounced at arbitrary regions. Notably, this effect was particularly evident at the NSA=12 level. The activation signal amplitude change of $B_1$ phase in the left motor cortex were as follows; NSA=12: $0.00016 \pm 0.00023$ , NSA=24: $0.00017 \pm 0.00012$ , NSA=48: $0.00019 \pm 0.00010$ , NSA=72: $0.00024 \pm 0.00010$ , NSA=144: $0.00018 \pm 0.00007$ , NSA=168: $0.00018 \pm 0.00006$, and NSA=192: $0.00021 \pm 0.00005$ [radian].

The activation maps of the reconstructed conductivity under various NSA levels are presented in Fig. 3B. Notably, with signal-averaging at NSA=24 and above, significant negatively correlated activation regions dominantly surrounded the left motor cortex region. In the extracted conductivity signals in the activation regions, noise fluctuations were still pronounced, similar to what was observed in the $B_1$ phase. The activation signal amplitude of conductivity in the left motor cortex were as follows; NSA=12: $-0.0010 \pm 0.0010$, NSA=24: $-0.0010 \pm 0.0008$, NSA=48: $-0.0013 \pm 0.0007$, NSA=72: $-0.0020 \pm 0.0007$, NSA=144: $-0.0018 \pm 0.0006$, NSA=168: $-0.0018 \pm 0.0006$, NSA=192: $-0.0025 \pm 0.0006$ [S/m]. Compared to the reference signal (NSA=192), as NSA level decreased (i.e., noise level increased), the reconstruction error became increasingly characterized by the quantitative underestimation of the conductivity amplitude change. Moreover, unlike the activation signal from the $B_1$ phase, the activation signal from the reconstructed conductivity exhibited slow variations before and after the onset of the finger-tapping stimulation due to the use of a large reconstruction kernel and the spatial denoising filter. To address the potential influence of regions without $B_1$ phase activation due to the use of a large reconstruction kernel, the conductivity reconstruction algorithm was applied specifically to the intersection area between the regions of interest (ROIs) of the activation maps in fMRI and $B_1$ phase (Fig. 3B: Intersection ROI Reconstruction; orange highlighted outline). In this intersection region, the observed conductivity signal demonstrated immediate changes in response to the stimulus similar to the $B_1$ phase signal observations.

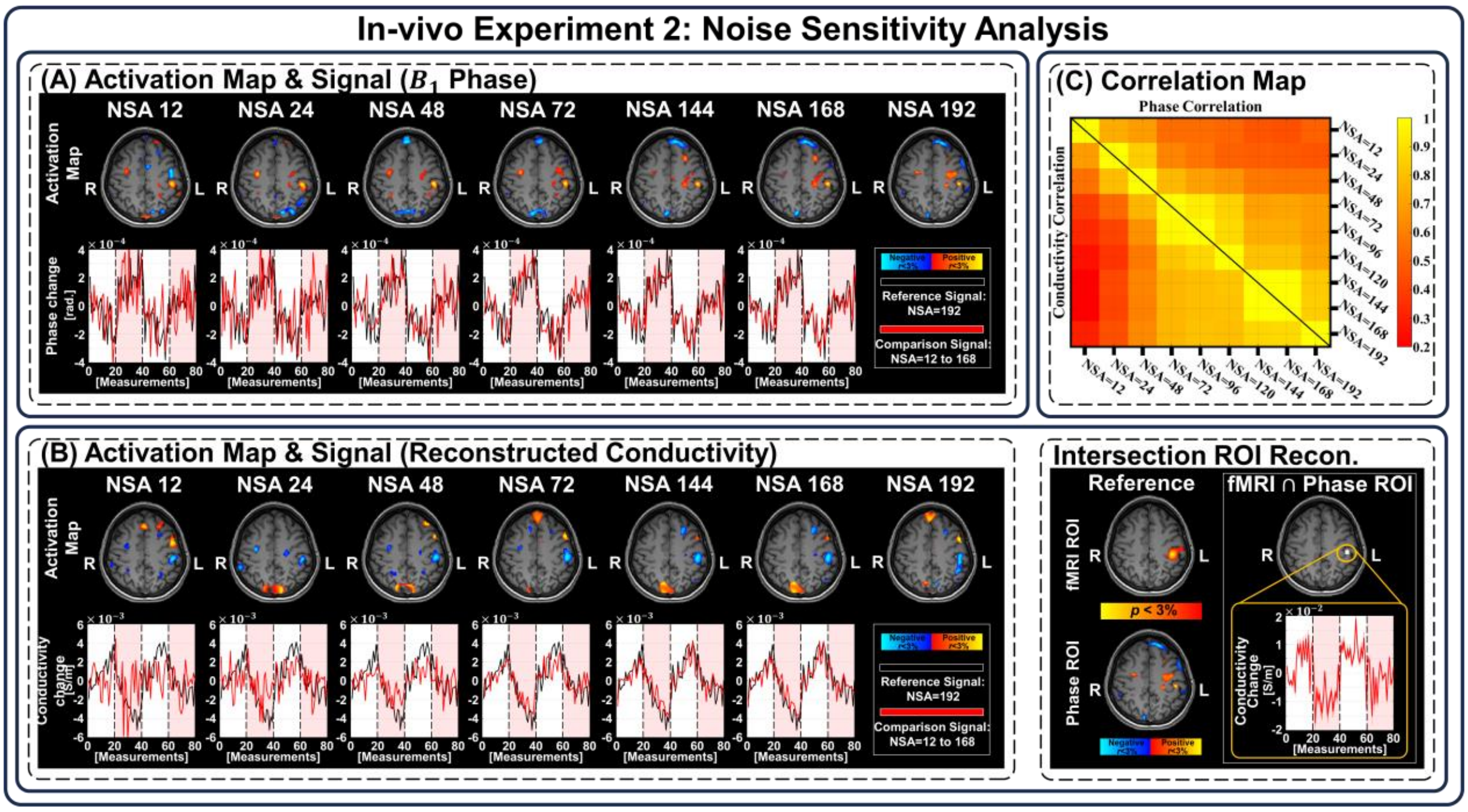

**Fig. 3|** In-vivo experiment 2: Noise sensitivity analysis through signal-averaging for $B_1$ phase and reconstructed conductivity. **A** Activation maps of $B_1$ phase and corresponding extracted mean signal (red) for various NSA levels (NSA=12, 24, 48, 72, 144, 168, and 192) in the left motor cortex. The $B_1$ phase signal from NSA=192 (black) was utilized for comparison. **B** Activation maps of reconstructed conductivity and corresponding extracted mean signal (red) for various NSA levels in the left motor cortex. The conductivity signal with NSA=192 (black) was utilized for comparison. Additionally, the conductivity was observed in the intersection region of the bSSFP fMRI and $B_1$ phase activation maps (NSA=192) to address the potential influence of regions without $B_1$ phase activation due to the use of a large reconstruction kernel. The conductivity temporal signal (red) shows sharp transitions. **C** Correlation maps of $B_1$ phase and conductivity across different NSA levels.

Noise sensitivity for the $B_1$ phase and reconstructed conductivity was evident in the correlation map across various NSA levels (Fig. 3C). Overall, the conductivity signals exhibited lower correlation values than the $B_1$ phase signals, indicating its susceptibility to noise from the reconstruction process. As shown in the signal observations in Fig. 3A and B, the correlation values for both $B_1$ phase and conductivity signals remained above 0.7 for NSA levels of 72 and higher. However, below this threshold, the correlation decreased sharply. Specifically, the correlation value between the $B_1$ phase signals was 0.46 when comparing NSA=16 and NSA=192, while the correlation between the conductivity signals was 0.31.

## Simulation-based Analysis: Investigating the Relationship between $B_1$ phase and Reconstructed Conductivity

In this experiment, we employed simulations to replicate in-vivo fMRI experiments by altering conductivity values in specific regions of both cylinder[29] and human brain[30,31] (Duke) phantoms. The primary objective was to investigate the relationship between $B_1$ phase and reconstructed conductivity using simulation models devoid of in-vivo spontaneous perturbations and motion. Additionally, we aimed to assess the signal sensitivity during $B_1$ phase and conductivity changes under various conditions. For the simulation experiments using the phase-based EPT algorithm, we employed the Poly-Fit method (with a default kernel size of 17×17). Details of simulation settings are summarized in the Methods Section.

We first investigated the relationship between $B_1$ phase and reconstructed conductivity with the cylinder phantom in the simulation experiment 1 (Fig. 4A). Due to the uncertainty in the relationship between changes in $B_1$ phase and conductivity, we examined six simulation models with different conductivity based activation models: three with increasing conductivity (Model 1: $\sigma_{GM} + 0.06$ [S/m], Model 2: $\sigma_{GM} + 0.04$ [S/m], Model 3: $\sigma_{GM} + 0.02$ [S/m]) and three with decreasing conductivity (Model 4: $\sigma_{GM} - 0.02$ [S/m], Model 5: $\sigma_{GM} - 0.04$ [S/m], and Model 6: $\sigma_{GM} - 0.06$ [S/m]) (Fig. 4B). In the simulated activation region, under noiseless conditions, higher conductivity amplitude change level resulted in a steeper phase curve (Fig. 4B: purple line in phase graph), while lower conductivity amplitude changes produced a more gradual curve (Fig. 4B: green line in phase graph). During this $B_1$ phase change, the steeper $B_1$ phase curve corresponded to a decrease in $B_1$ phase within the activation region, whereas the more gradual curve was associated with an increase in $B_1$ phase. Additionally, the activation amplitude change trend of the reconstructed conductivity closely matched that of the ground-truth conductivity (GTC). An inverse correlation was consistently seen between $B_1$ phase and reconstructed conductivity, regardless of simulated amplitudes of conductivity (Fig. 4B: phase graph and reconstructed conductivity graph). In the activation map processed under noise conditions similar to in-vivo level, this inverse relationship between $B_1$ phase and conductivity persisted (Fig. 4C). Furthermore, spurious activation signals due to noise occurred and could be seen in arbitrary regions where conductivity perturbation was not assigned.

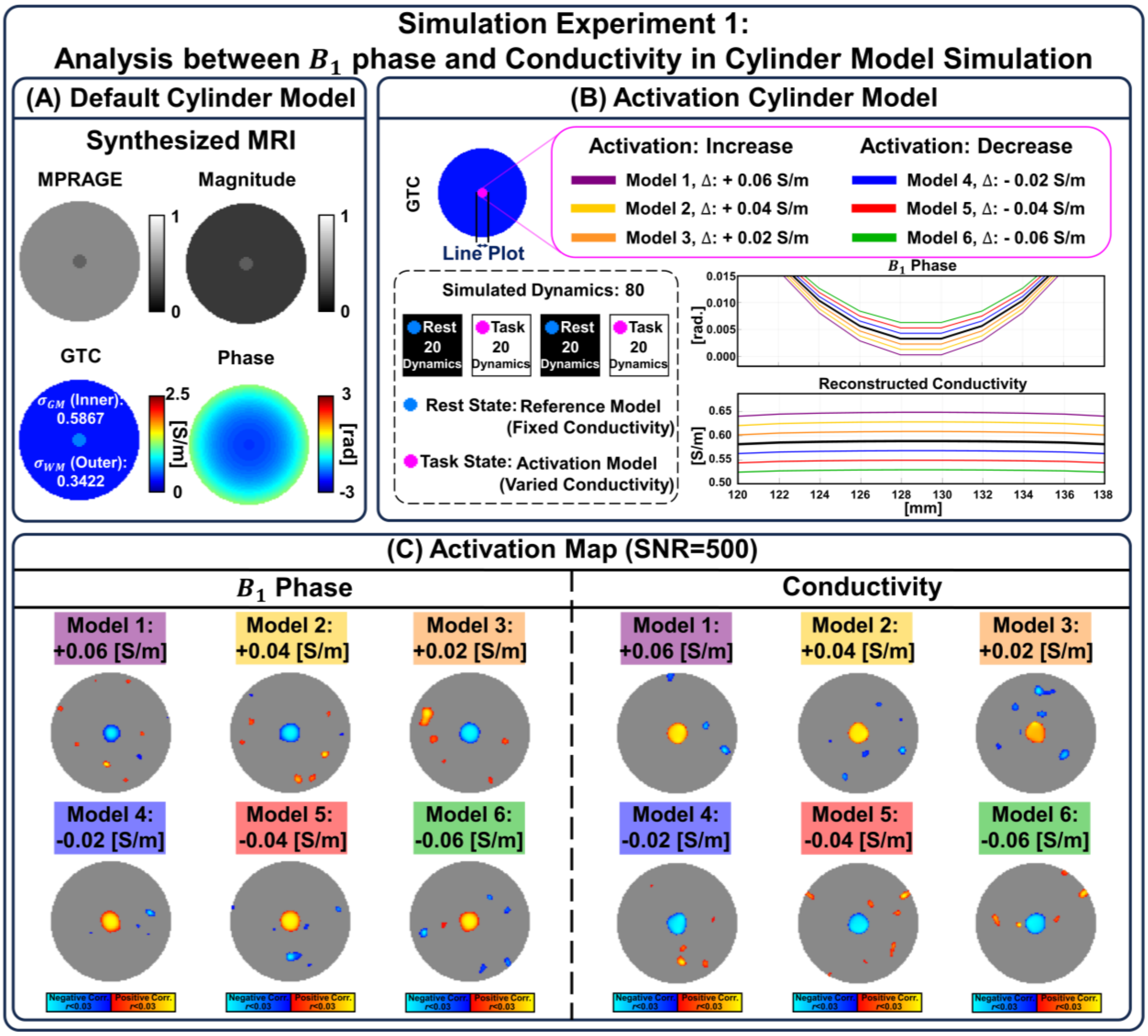

**Fig. 4|** Simulation experiment 1: Analysis between $B_1$ phase and conductivity in cylinder model simulation. **A** Default model for cylinder simulation experiment. **B** Activation model for the cylinder model simulation and line plot of activation regions in phase, along with corresponding reconstructed conductivity maps (noiseless conditions). Six scenarios were simulated for the activated conductivity models; Model 1: $\sigma_{GM} + 0.06$ S/m, Model 2: $\sigma_{GM} + 0.04$ S/m, Model 3: $\sigma_{GM} + 0.02$ S/m, Model 4: $\sigma_{GM} - 0.02$ S/m, Model 5: $\sigma_{GM} - 0.04$ S/m, and Model 6: $\sigma_{GM} - 0.06$ S/m. **C** Activation maps for $B_1$ phase and conductivity in various conductivity modified activation models (*r*<0.03).

Next, we conducted the simulation experiment 2 using the heterogeneous brain phantom that mimics the complex structure of in-vivo conditions. The goal of this experiment was to determine whether the relationship between $B_1$ phase and reconstructed conductivity observed in the cylinder phantom experiment would similarly manifest in the complex brain structure. Based on the results from the in-vivo and cylinder phantom experiments, which demonstrated a trend of increased $B_1$ phase and decreased conductivity, the initial brain simulation experiments focused on models designed to simulate decreasing conductivity during activation. As shown in Fig. 5A, the simulated brain models include: Model 7 ($\sigma_{GM} - 0.02$ S/m), Model 8 ($\sigma_{GM} - 0.04$ S/m), and Model 9 ($\sigma_{GM} - 0.06$ S/m). Subsequently, we conducted sensitivity experiments under other conditions, including various noise levels, noise repeatability, and simultaneous activations of both positive and negative conductivity changes in the motor cortex region.

First, we observed the activation signal amplitude of the models in the activation regions for three scenarios (Fig. 5B). In this observation, we used the Poly-Fit method for the conductivity reconstructions under the same condition as the in-vivo experiments 1. Based on the reference in-vivo mean amplitude signals from the in-vivo experiment 1, the activation signal amplitude of the model with conductivity change set to $-0.04$ [S/m] (Model 8) exhibited a trend similar to the in-vivo experiment 1 (Fig. 2 and Fig. 5B: Mean (STD) for Activation Region); for in-vivo, $B_1$ phase: $0.00030$ [radian] and conductivity: $-0.00192$ [S/m]; for brain simulation, $B_1$ phase: $0.00032$ [radian] and conductivity: $-0.00174$ [S/m]. Compared to the in-vivo signal changes, the simulation experiments showed a tendency for the $B_1$ phase to be slightly overestimated, while the conductivity tended to be underestimated. Note that even though the conductivity value is in [S/m] unit, it does not reflect the actual value change due to spatial filtering (see Discussion section). Using Model 8, we examined the activation maps using the $B_1$ phase difference map (between simulated default and activation models) as a reference. In the heterogeneous brain phantom, even though only decreased conductivity change ($\sigma_{GM} - 0.04$) was assigned to specific regions of the motor cortex, positively correlated changes in $B_1$ phase were dominant with some negatively correlated changes accompanying (Fig. 5B: Phase Activation Map for Model 8). For the activation maps resulting from the reconstructed conductivity, we investigated various phase-based EPT algorithms, including phase-based Laplacian EPT[27], convection-reaction (cr) EPT[32], Poly-Fit[28], and segmentation-based integral EPT[33] (integral-based EPT), with different kernel sizes. At realistic SNR levels (SNR=500), the use of small kernels (e.g., below 9×9 kernel size) was either inappropriate or resulted in spurious activations around the simulated activation region (Fig. 5B and Fig. S2: Laplacian EPT and cr-EPT). In contrast, cr-EPT, Poly-Fit and integral-based methods, which utilized sufficiently large kernels (e.g., above 11×11 kernel size), displayed activation maps with the negatively correlated signals around the corresponding activation regions. However, in the case of the integral-based method, the segmented regions in the WM around the motor cortex predominantly included negatively correlated $B_1$ phase distributions, resulting in the creation of spurious positively correlated activation regions in the conductivity activation maps, regardless of reconstruction kernel size (Fig. S2: Integral-based EPT).

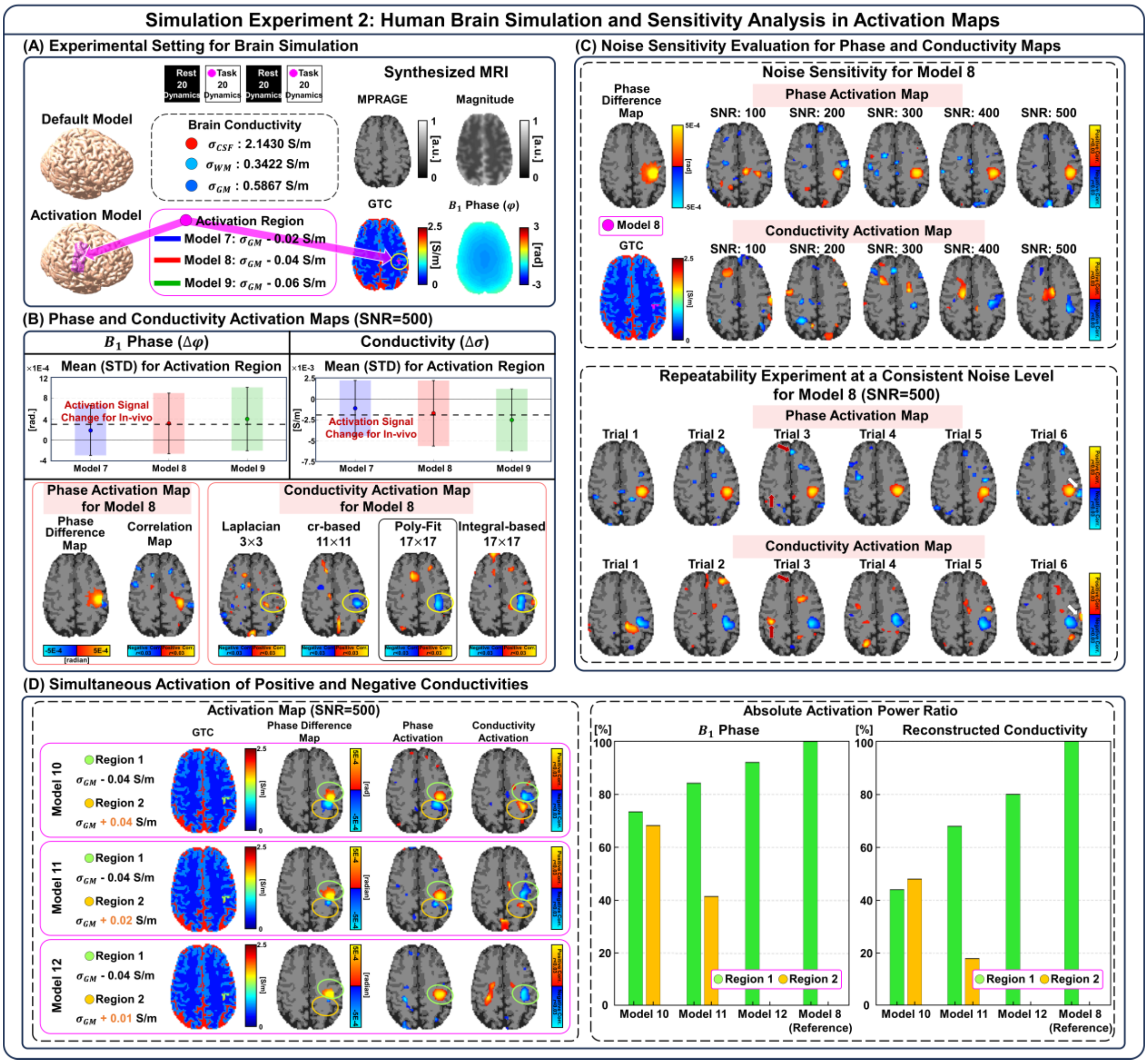

**Fig. 5|** Simulation experiment 2: Human brain simulation and sensitivity analysis in activation maps. **A** Default and activation models (Model 7 ($\sigma_{GM} - 0.02$ S/m), Model 8 ($\sigma_{GM} - 0.04$ S/m), and Model 9 ($\sigma_{GM} - 0.06$ S/m)) for human brain simulation experiment. The left motor cortex region was selected for activation simulation (purple arrow). **B** Estimated signal amplitude in activation regions for $B_1$ phase ($\Delta\varphi$) and conductivity ($\Delta\sigma$). In the bar graphs for observing the mean activation signal changes, the mean signal change from in-vivo experiment (Fig. 2) was used as the reference. Additionally, various phase-based reconstruction methods were applied and compared based on the Model 8 (phase-based Laplacian EPT, cr-EPT, Poly-Fit, and integral-based EPT). For the conductivity reconstruction, we primarily employed the Poly-Fit method (17×17). **C** Noise sensitivity experiment using Model 8. At SNR=500, consistent activation using the conductivity map in the left motor region can be seen. **D** Experiment on the simultaneous activation of positive and negative conductivity activations. When the sensitivity to opposing activations in simultaneous activation simulations falls below 25%, it becomes practically impossible to distinguish these activations in both the $B_1$ phase and the reconstructed conductivity maps.

Subsequently, we investigated the sensitivity to noise with various noise levels (SNR=100 to 500 with 100 steps) (Fig. 5C: Noise Sensitivity for Model 8). In the activation map from the $B_1$ phase, positively correlated activation regions were detected around the simulated activation regions at all SNR levels. However, at SNR levels below 200, the activation map from the $B_1$ phase exhibited false-positive activations (negatively correlated activation) near the motor cortex. This trend was also observed in the reconstructed conductivity at SNR levels below 200. Due to the characteristics of the phase-based EPT algorithm, amplification in noise led to spurious activations in various arbitrary regions. Especially, at SNR levels of 100 and 200, it was indicated that activation from the corresponding motor cortex cannot be detected in the reconstructed conductivity map. At SNR levels above 300, negatively correlated activation regions were seen in the corresponding simulated activation areas of the motor cortex while a slight shift in the activation region was observed after the phase-based EPT reconstruction process.

Next, we conducted a repeatability experiment at a consistent noise level (SNR=500) (Fig. 5C: Repeatability Experiment at a Consistent Noise Level for Model 8). In the activation map from the $B_1$ phase, positively correlated activation was consistently detected in the simulated activation regions. However, negatively correlated activation regions caused by noise also appeared in various arbitrary locations across all repeated trial models. Notably, in the Trial 6 model, these spurious activations even emerged around the simulated activation regions (Fig. 5C, white arrows). Furthermore, in the Trial 3 model, spurious activation was observed in the contralateral region of the reconstructed conductivity map, even when no activation correlation was detected in the $B_1$ phase (Fig. 5C, red arrows). These observations indicate that noise in the $B_1$ phase could easily produce spurious activation signals in the reconstructed conductivity maps, sometimes manifesting as positively correlated activation around the activation region.

In the subsequent experiment, we explored scenarios involving simultaneous activation producing both positive and negative conductivity changes within the activation region (Fig. 5D). For this scenario, we utilized Model 8 as a baseline, fixing a region with a conductivity decrease of $-0.04$ [S/m], and positive conductivity changes were then assigned in nearby regions with the activation conductivity amplitude changes of 0.04, 0.02, and 0.01 [S/m], respectively corresponding to relative conductivity increase of 100 [%], 50 [%], and 25 [%]. As shown in Fig. 5D: Model 10, when the amplitudes of the positive and negative conductivity changes were equal, distinct activation regions were observed in both the activation maps of the $B_1$ phase and the reconstructed conductivity. In both the 100 [%] and 50 [%] amplitude change cases, positively and negatively correlated activation regions remained distinguishable in both the $B_1$ phase and the reconstructed conductivity (Models 10 and 11). However, at opposing 25 [%] conductivity change (Model 12), the activation map from the $B_1$ phase exhibited only positively correlated activation, which also led to negatively correlated activation emerging as the prevailing feature in the reconstructed conductivity. Using the activation signal from Model 8 as a reference, we analyzed the relative activation power ratio within both the positively and negatively correlated activation regions in the reconstructed conductivity for each model (Fig. 5D: Absolute Activation Ratio). The presence of oppositely correlated activation signals resulted in a relative reduction in the observed activation power. However, with a 25 [%] activation amplitude change, the relative activation power was not discernible in the opposing activation region. These results can provide information with regard to the level of conductivity change that can be simultaneously detected.

## In-vivo Visual Cortex Stimulation Response using fMRI, $B_1$ phase, and Conductivity

To observe whether $B_1$ phase and conductivity changes could also occur in other tasks, we applied the same process to checker-board visual stimulus and observed the corresponding $B_1$ phase and reconstructed conductivity changes. In the in-vivo experiment 3, we performed GE-EPI scans (NSA=1) and bSSFP scans (NSA=24) during visual stimulation. For the in-vivo experiment 4, we conducted observations with high SNR levels, achieved through signal-averaging (Total NSA=192). For the phase-based EPT algorithm, we employed the Poly-Fit method for both in-vivo experiments 3 and 4 (kernel size=11×11). Additionally, we investigated the effects of the larger reconstruction kernel (kernel size=17×17). Details of the experimental setting are described in the Methods section.

In the in-vivo experiment 3, we first observed the activation maps according to GE-EPI (Fig. S3A) and bSSFP (Fig. 6A) ($p$<0.03). Similar to the observations in the motor cortex experiments, while the fMRI activation regions were prominent in the visual cortex for both modalities, the specific areas of activation varied subtly depending on the imaging sequence. We analyzed the mean signal changes in the visual cortex for all volunteers ($n$=4). For all volunteers, the activation signal amplitude was prominent in the visual cortex; GE-EPI: $0.0093 \pm 0.0032$ [a.u.] versus bSSFP: $0.0012 \pm 0.0002$ [a.u.] ($p$<0.03). Same as observations in motor cortex stimuli experiments, GE-EPI demonstrated increased fMRI signal sensitivity compared to bSSFP; GE-EPI: $1.0442 \pm 0.3778$ [%] versus bSSFP: $0.3021 \pm 0.0815$ [%] ($p$<0.03). For both GE-EPI and bSSFP, the signal change amplitude in response to visual stimuli was relatively smaller compared to the finger-tapping stimuli. Secondly, we examined whether the visual stimulus led to localized changes in $B_1$ phase distributions around the visual cortex. The activation maps from $B_1$ phase information are presented in Fig. 6B. The activation map from the $B_1$ phase did not coincide exactly with the bSSFP fMRI map but showed a dominant presence of positively correlated signal changes in the surrounding fMRI areas. In addition, unlike the $B_1$ phase changes observed in response to motor cortex stimuli, visual cortex stimuli resulted in activation regions appearing in multiple or broader phase activation areas surrounding the fMRI region (Fig. 2A versus Fig. 6B). The activation signal amplitude from $B_1$ phase in visual cortex regions for all volunteers ($n$=4) was $0.00022 \pm 0.00017$ [radian]. The percentile change in response to visual cortex stimuli was lower compared to motor cortex: visual cortex, $0.0398 \pm 0.0188$ [%]; motor cortex, $0.0547 \pm 0.0306$ [%]. Considering the characteristics of the $B_1$ phase change distribution, we applied the Poly-Fit algorithm with a smaller kernel size ($11 \times 11$) to analyze the conductivity changes (Fig. S3B). The conductivity-based activation maps consistently showed that the conductivity changes were predominantly negatively correlated (Fig. 6C). The activation signal amplitude of conductivity was observed in the visual cortex region ($n$=4): $-0.0019 \pm 0.0012$ [S/m]. Compared to the conductivity reconstruction results in the motor cortex, the use of a relatively smaller reconstruction kernel resulted in a higher amplitude of conductivity changes in response to visual cortex stimuli: $-0.0398 \pm 0.0188$ [%]. In addition, compared to the motor cortex stimuli experiments, both $B_1$ phase and conductivity activation signals in the visual cortex exhibited relatively lower amplitudes but demonstrated similar activation correlation trends.

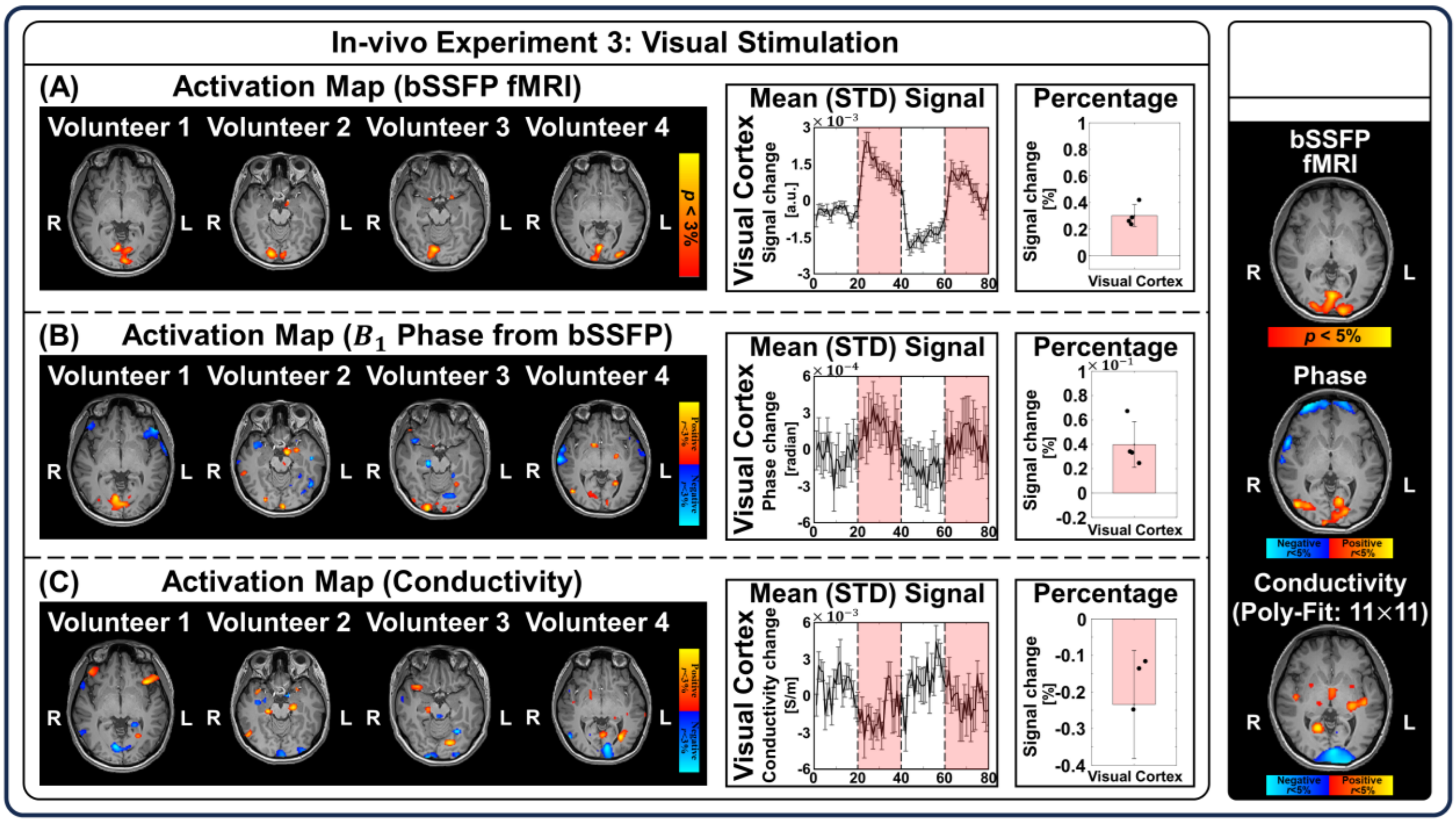

**Fig. 6|** In-vivo experiments 3 and 4 for visual stimuli. In-vivo experiment 3 ($n$=4): bSSFP fMRI, $B_1$ phase and reconstructed conductivity activation maps with mean (STD) signal and activation percentage in target regions. **A** Activation maps from bSSFP magnitude signal. The activation regions were detected at a significance level of $p$<0.03. Mean (STD) temporal-series signal derived based on the visual cortex (activation) regions. **B** $B_1$ phase activation map obtained from bSSFP. Both positive and negative correlation coefficients were considered as significant, utilizing a two-tailed significance threshold of $r$<0.03. Mean (STD) temporal signals derived from visual cortex regions. **C** Reconstructed conductivity activation map obtained from the bSSFP phase. The Poly-Fit was utilized for the phase-based reconstructions (11×11). Mean (STD) temporal signals derived from the visual cortex regions. **D** In-vivo experiment 4: Visual stimuli experiment conducted on a volunteer with high signal-averaging (NSA = 192). Activation maps from bSSFP fMRI, $B_1$ phase, and reconstructed conductivity. bSSFP fMRI activation signals were detected at a significance level of $p$<0.05. For the $B_1$ phase and conductivity maps, both positive and negative correlation coefficients were considered as significant, utilizing a two-tailed significance threshold of $r$<0.05.

In the in-vivo experiment 4 using high SNR acquisition, we first examined the activation maps from GE-EPI and bSSFP (Fig. S3B and Fig. 6D). The prominent activation regions varied subtly depending on the imaging sequence. Similar to the observations in the in-vivo experiment 3, significant $B_1$ phase changes were distributed across various and broadened areas near the visual cortex ($r$<0.05). This distribution characteristic of the $B_1$ phase changes can influence the reconstructed conductivity, depending on the reconstruction kernel size used (Fig. 6D: Poly-Fit 11×11 and Fig. S3B: Poly-Fit 17×17). Noticeably, using a relatively small kernel (Fig. 6D) revealed negatively correlated conductivity changes around the fMRI activation region, while a larger kernel (Fig. S3B: middle row) showed positively correlated conductivity changes. To exclude the influence of regions without $B_1$ phase activation due to the large kernel, the conductivity reconstruction algorithm was applied to the $B_1$ phase within the intersection area between the ROIs of fMRI and $B_1$ phase activations (Fig. S3B: ROI-based Reconstruction). Within this intersection ROI, the conductivity exhibited a negatively correlated change, consistent with the results from the finger-tapping experiments.

# Discussion

In this study, we demonstrated in-vivo electrical conductivity changes in the brain during fMRI experiments. We observed that negative correlation with activation was predominantly observed in the conductivity values while positive correlation with activation was dominant in the $B_1$ phase. Within the regions of activation, as is well known, GE-EPI based BOLD showed increased signal intensity with task activation due to increased blood flow and a resulting increase in the oxyhemoglobin/deoxyhemoglobin ratio, which makes the environment more diamagnetic. Interestingly, for conductivity, the changes are negatively correlated with task activation. Based on the simulation experiments, the functional activity related quantitative change in conductivity is expected to be approximately $-0.04$ S/m. Additionally, the simulation experiments demonstrated that spurious false-positive signals could emerge outside the simulated activation regions at certain noise level. We employed signal-averaging with a bSSFP sequence as a strategy to suppress noise, and our study demonstrates that conductivity changes can be observed with repeatability using clinical 3T MRI scanners. To date, previous studies[19,20,21,22,23,24] have reported mixed findings regarding conductivity changes during fMRI; however, this study for the first time provides robust demonstration of how conductivity in-vivo varies in the context of activation. Finally, the controlled simulation studies confirm the in-vivo observations and highlight that reconstruction methods can influence the observation of conductivity.

The source of the observed changes in conductivity remains to be fully determined. It is known that differences in imaging modalities can be influenced by different aspects during functioning such as vessel size, orientation, and blood volume fraction[7]. While increases in ion-concentrations and blood flow are well-known to be major contributors to increased conductivity[34,35,36,37], several reports have suggested that changes in blood components and/or oxygenation levels can lead to a decrease in conductivity. Firstly, it was demonstrated that red blood cell concentration can influence resistance, with higher red cell density leading to higher resistance[38] (i.e., lower conductivity). Secondly, it was reported that hemoglobin concentration itself in the blood can affect conductivity[39]. Conductivity decreased as hemoglobin density increased compared to the initial count, although the relationship did not follow a linear trend. Thirdly, it was observed that in blood samples, conductivity increases as oxygen saturation decreases over time[40], which indirectly indicates that the concentration of oxygen level in the blood can affect conductivity. Another source of conductivity decrease during activation could be the relative reduction in CSF volume[41]. These findings collectively suggest that the observed decrease in conductivity during neural activation could be attributed to a complex interplay of the aforementioned factors.

'Functional conductivity' can offer a unique perspective on brain activity that complements established techniques such as BOLD, vascular space occupancy[42,43] (VASO), perfusion[44,45], and cerebral metabolic rate of oxygen[46,47] ($CMRO_2$) measurements. While BOLD fMRI primarily reflects hemodynamic changes associated with neural activity, 'functional conductivity' directly measures changes in the electrical properties of brain tissue, potentially revealing aspects of neural activity not captured by hemodynamic-based methods. VASO, perfusion, and $CMRO_2$ primarily provide information about hemodynamic processes, such as changes in cerebral blood volume, blood flow, and metabolic rate of oxygen. In contrast, 'functional conductivity' offers a different perspective by revealing how these hemodynamic changes influence the brain's electrical properties. Whereas the former techniques focus on the dynamics of blood and metabolism, 'functional conductivity' can provide insights into how these physiological changes affect the overall electrical environment of brain tissue and reveal how metabolic processes impact tissue conductivity, potentially enhancing our understanding of the interplay between vascular factors and electrical properties. We note that conductivity can be simultaneously acquired with BOLD when using a multi-echo GE-EPI sequence, which may lead to increased sensitivity of functional activity[24,48,49].

Despite its potential, there are several challenges that must be addressed before functional conductivity can be practically implemented. In our in-vivo experiments, despite employing a signal averaging strategy (NSA=24), the $B_1$ phase and conductivity activation signals still remained susceptible to noise and spontaneous activity. This noise susceptibility was evident when comparing activation signals across varying NSA levels. To address these challenges, we applied a correlation analysis approach to enhance the extraction of activation maps from $B_1$ phase and reconstructed conductivity. The correlation analysis contributed to mitigating the impact of noise and spontaneous activity, thereby enhancing the reliability of the activation maps derived from the $B_1$ phase and conductivity data. The use of the general linear model (GLM) method would have led to heightened sensitivity to noise complicating the reliable extraction of activation maps as demonstrated in Fig. S4.

For the in-vivo experiment, we also performed a SE-EPI[50,51] scan on the same slice (Fig. S1). In the activation maps of the SE-EPI phase, we noted the occurrence of both positive and negative correlated activations around the activation regions, or in some instances, a complete lack of activation correlation. This could be due to the lower SNR efficiency of SE-EPI compared to bSSFP[52]. Nevertheless, for the activation maps from the reconstructed conductivity from SE-EPI, similar negative correlations were mainly predominant within the activation regions; however, the strength of these correlations was generally weak. In certain cases, only positive correlations were observed.

While our observations revealed that positively correlated activation is dominant in the $B_1$ phase and negatively correlated activation prevails in the reconstructed conductivity, as demonstrated in the brain simulation studies, both positive phase changes and minor negative phase changes can occur (Fig. 5B). This finding raises the challenge that it is difficult to distinguish between these two cases using current phase-based EPT algorithms (Fig. 5D) when opposing activation is simultaneously present. This occurs because of the non-local properties of phase; $B_1$ phase changes are not confined to only the activation region but also influence the surrounding tissues. Furthermore, in the phase-based EPT reconstruction, the resulting activation map can vary depending on the tissue size[53] and reconstruction kernel size[54]. In the case of the motor cortex region (default kernel size=17×17), $B_1$ phase revealed activation concentrated within specific areas of the motor cortex, making it relatively unaffected by the reconstruction kernel size. On the other hand, in the case of the visual cortex (default kernel size=11×11), $B_1$ phase-based activation occurred across various areas near the visual cortex. Using a large kernel led to the appearance of positively correlated conductivity activation near the visual cortex (Fig. S3B: Poly-Fit result with 17×17). These findings underscore the importance of kernel size selection in conductivity reconstruction, as it can influence the result of activation maps. Additionally, the simulations mimicking the in-vivo experimental environment estimated that the functional signal-related conductivity change in motor cortex activation would be around $-0.04$ S/m. However, in in-vivo applications, the use of a large reconstruction kernel, intended to suppress noise, hinders the accurate tracking of actual quantitative conductivity changes. Future work should explore the optimization of kernel size parameters to further enhance the reliability of conductivity reconstructions in various fMRI settings.

The conductivity-based activation region tends to shift compared to the $B_1$ phase-based activation region. This occurs because the phase-based reconstruction process relies on the surrounding $B_1$ phase curvature information, incorporating non-activated regions into the reconstruction process due to the use of a relatively large kernel. Furthermore, the phase-based reconstruction algorithm is prone to generating spurious false-positive signals, which can appear in the reconstructed conductivity even when no significant correlated activation region is present in the $B_1$ phase. To avoid the observation of these spurious signals, it is essential to compare activation regions observed in both fMRI and $B_1$ phase activation maps with the activation map of the conductivity.

Despite these limitations, there are several potential avenues for advancing the application of functional conductivity measurements. We note that the EPT technique used in this study relies on measurements at the Larmor frequency (i.e., 128 MHz at 3T). Consequently, the observed conductivity values correspond to this specific frequency, reflecting contributions from both intracellular and extracellular components. While direct measurement of intracellular conductivity changes could potentially provide a deeper understanding of the overall changes observed, such measurements are currently not feasible in a non-invasive manner[55]. Studies that explore the separation of intracellular and extracellular conductivity, such as those based on two-compartment models, have been suggested and may provide usefulness for further insights[56].

In this study, we focused on conductivity due to its relative ease of measurement. Due to the relatively longer scan times and susceptibility to noise associated with $B_1^+$ magnitude ($|B_1^+|$) mapping[57,58,59], much of the research to date has primarily focused on phase-based conductivity reconstructions. However, it is important to note that tissue electrical properties influenced by the RF field are characterized by complex permittivity ($\epsilon^* = \epsilon - j(\sigma/\omega)$). Therefore, future research could also investigate the functional related signal changes to permittivity[27], which would be dominated by the $|B_1^+|$. Understanding these intricate relationships will not only enhance the understanding of functional MR imaging source but also provide deeper insights into the complex interplay between electrical properties and neural activity.

# Methods

## Experiment Participants

We recruited 11 healthy volunteers (6 males and 5 females; mean age 25 years, range 22–32 years) for the finger-tapping and visual stimulations in the fMRI experiment. The scans were performed under the approval of the local institutional review board (Yonsei University, IRB, no: 7001988-202302-HR-1784-02). All participants gave written informed consent before taking part in the study. They were all right-handed, neurologically healthy, and had either normal vision or vision corrected to normal.

## MRI Experimental Setting

We investigated functional MRI, $B_1$ phase, and conductivity changes induced by neural activity in healthy volunteers. To observe these activation changes, we employed the following experimental setting. Utilizing an alternating block design (with four alternating blocks of resting and task states, each block comprising 20 dynamics), our investigation included right-hand finger-tapping (total subject: $n$=11) and visual stimulation (total subject: $n$=5) experiments.

Volunteers were scanned on a 3T MRI system (MAGNETOM Vida, Siemens Healthineers) using 64 channel head coil. For each volunteer, 3D multi-slice GE-EPI, 2D SE-EPI, and 2D bSFFP sequences were employed for the functional imaging with the following scan parameters; GE-EPI: TR/TE = 3000/30 [ms], flip-angle = 90°, resolution = 3.8 [mm] × 3.8 [mm], slice thickness = 5 [mm], number of slices = 33, acceleration factor = 2, NSA = 1, number of dynamics = 90, bandwidth = 2894 [Hz], and total scan time = 4:39; SE-EPI: TR/TE = 3000/60 [ms], flip-angle = 90˚, resolution = 3.8 [mm] × 3.8 [mm], slice thickness = 5 [mm], number of slices = 1, NSA = 1, number of dynamics = 90, bandwidth = 2894 [Hz], and total scan time = 4:36; bSSFP: TR/TE = 4/2 [ms], flip-angle = 30˚, resolution = 3.8 [mm] × 3.8 [mm], slice thickness = 5 [mm], number of slice = 1, NSA = 12, number of dynamics = 90, bandwidth = 514 [Hz], and total scan time = 4:40. For all functional MRI acquisitions, the first 10 dynamics were discarded from the analysis. Additionally, for brain anatomy scan, 3D magnetization prepared rapid acquisition gradient echo (MPRAGE) sequence was utilized with following scan parameters: TR/TE = 2200/2.91 [ms], flip-angle = 8˚, resolution=1.0 [mm] × 1.0 [mm] × 1.0 [mm], the number of slices = 208, acceleration factor = 2, NSA = 1, bandwidth = 180 [Hz] and total scan time = 5:27. For the GE-EPI, SE-EPI, and bSSFP dataset, the image resolution was interpolated to 1.9 [mm] × 1.9 [mm].

The SNR of the MR contrast image was calculated using the following equation:

$$SNR = S_{WM}/((1/\sqrt{2-\pi/2}) \cdot \tau_{noise}) \quad (1)$$

where $S_{WM}$ is the mean signal intensity in a homogeneous WM region, and $\tau_{noise}$ is the standard deviation of the noise measured from the background that contains no MR signal intensity. The coefficient $(1/\sqrt{2-\pi/2})$ is applied to adjust for the conversion of MRI noise distribution from Gaussian to Rayleigh, accounting for the change in noise characteristics as the MRI data is transformed.

## MRI Scan Strategy for observing $B_1$ phase and conductivity changes

Initially, a 3D multi-slice GE-EPI scan was performed to identify a specific slice with pronounced BOLD activation. This selected slice was targeted in subsequent scans using SE-EPI and bSSFP sequences. To counteract the eddy current effects inherent in $B_1$ phase during MRI scans[60], each SE-EPI and bSSFP sequence was executed twice with opposite readout directions. Each scan sequence was designed to capture temporal dynamic in neural activity with a temporal-resolution of 3 seconds. A detailed MRI scan process is described for the representative finger-tapping task in Fig. 7.

In the in-vivo experiment 1, volunteers performed a right-hand finger-tapping task. The GE-EPI scan was performed with NSA=1, the SE-EPI scan with total NSA=2, including eddy current correction (AP: NSA=1 and PA: NSA=1), and bSSFP scan with NSA=24, including eddy current correction (AP: NSA=12 and PA: NSA=12). In the in-vivo experiment 2, a single volunteer performed a right-hand finger-tapping task. The GE-EPI scan was performed with NSA=1 and bSSFP scan with NSA=192, including eddy current correction (AP: NSA=96 and PA: NSA=96). In the in-vivo experiment 3, volunteers were subjected to checker-board visual stimulation. The GE-EPI scan was performed with NSA=1 and bSSFP scan with NSA=24, including eddy current correction (AP: NSA=12 and PA: NSA=12). In the in-vivo experiment 4, a single volunteer was subjected to checker-board visual stimulation. The GE-EPI scan was performed with NSA=1 and bSSFP scan with NSA=192, including eddy current correction (AP: NSA=96 and PA: NSA=96).

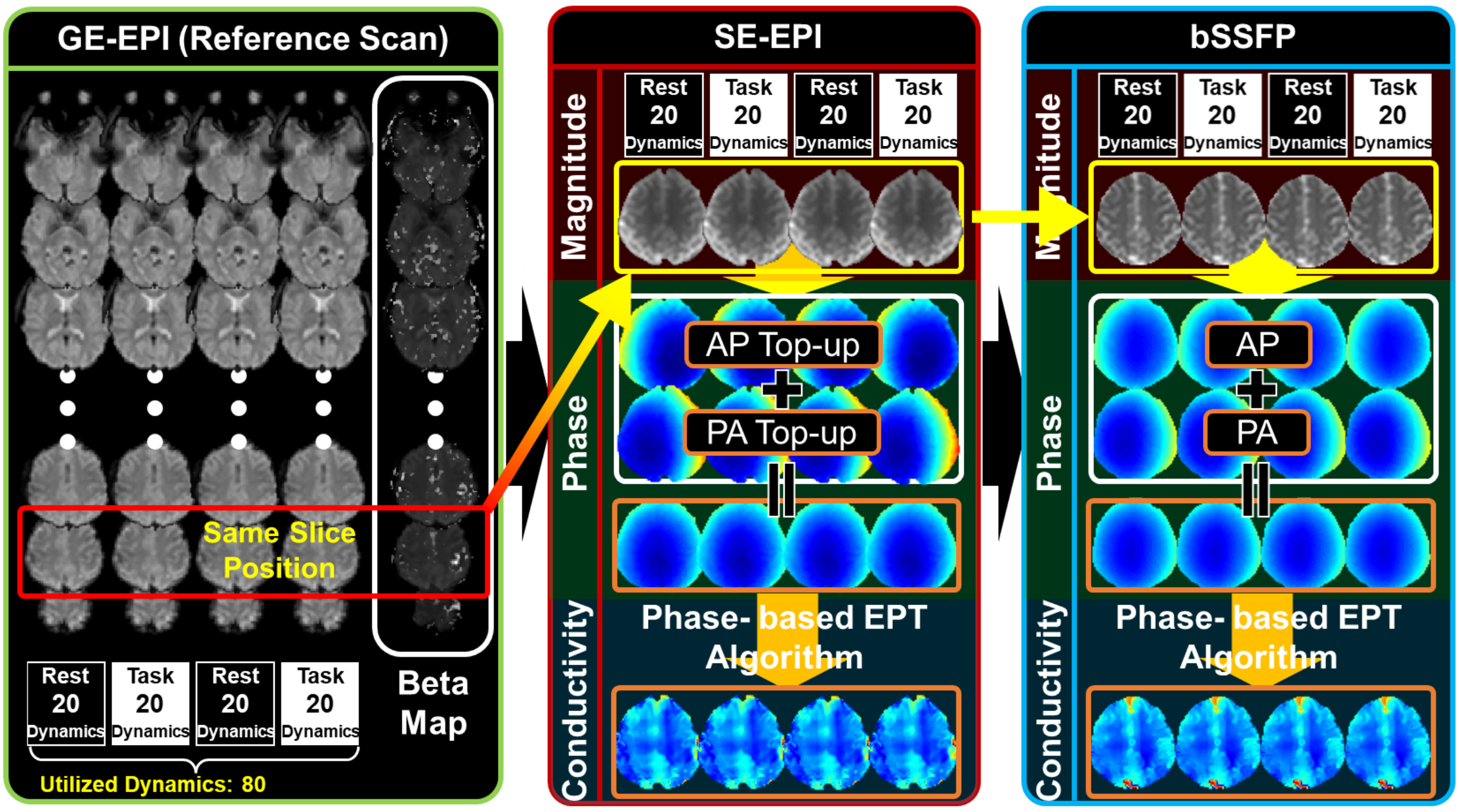

**Fig. 7|** MRI scan process. A multi-slice GE-EPI scan mapped the BOLD fMRI signal, identifying a slice with pronounced activation. This slice was targeted in subsequent SE-EPI and bSSFP scans. Eddy current effects were counteracted by executing the sequences twice with opposite readout directions. Phase-based EPT algorithm was used to reconstruct the conductivity from the $B_1$ phase data.

Based on the scanned MRI data, the SE-EPI and bSSFP images were processed as follows. For SE-EPI, the top-up method[61] (FSL software package[62]: http://fsl.fmrib.ox.ac.uk/fsl/fslwiki/FSL) was applied to each dataset scanned in both scan directions, and the results were combined to correct the eddy current effect. In the case of bSSFP, only eddy current correction was performed. Next, a linear phase shift correction was applied to the combined $B_1$ phase data using first-order polynomial coefficients to remove the residual phase shifts, to minimize any linear variations across the region of interest and the temporal series. During the registration process, to address the sparse structural information inherent in the $B_1$ phase information, a registration process was carried out using MR contrast information. The registration coefficients derived from MR contrast information were subsequently applied to the $B_1$ phase to mitigate motion-related artifacts. Then we applied the phase-based EPT algorithm[28] (see below) to the $B_1$ phase information to compute the conductivity.

To compute the activation map, a Gaussian filter with a full width at half maximum (FWHM) of 4.71 mm (standard deviation = 2) was applied to the MR magnitude, $B_1$ phase, and reconstructed conductivity information. Specifically, for the reconstructed conductivity of the visual stimulation data, a FWHM of 4.71 mm (standard deviation = 3.5) was applied due to the use of a relatively small reconstruction kernel (see Phase-based Conductivity Reconstruction section). Subsequently, we performed a first-order correction on the temporal-series signal of each spatial location. We note that no additional denoising filters were applied to the temporal-series signals in this study.

The BOLD map from GE-EPI and the fMRI maps from SE-EPI and bSSFP magnitude data were computed using GLM based on the Statistical Parametric Mapping[63] 12 (SPM12 software package: https://www.fil.ion.ucl.ac.uk/spm/software/spm12). Given the noise susceptibility and the ambiguity in defining activation regions for $B_1$ phase and conductivity, we employed a correlation analysis strategy based on the block design to extract the activation map. Both positive and negative correlation coefficients were computed to incorporate the uncertain activation in $B_1$ phase and conductivity.

## Phase-based Conductivity Reconstruction

In this study, we employed the phase-based weighted polynomial fitting reconstruction method[28]. The phase-based MR-EPT equation related the curvature of the measured $B_1$ phase ($\varphi$) to the tissue conductivity with piece-wise constant assumption ($\nabla\epsilon^* = 0$):

$$\sigma = \nabla^2\varphi/2\mu_0\omega \tag{2}$$

where $\sigma$ is measured conductivity, $\mu_0$ is the permeability of the free space, and $\omega$ is the Lamor frequency. Instead of forward computation of the Laplacian operator, the $B_1$ phase can be parameterized by the second polynomial coefficient.

$$\varphi = V\beta \tag{3}$$

Here $V$ is the so called the Vandermonde matrix, representing polynomial values at different kernel locations, and β are the polynomial coefficients (for 1, $x, y, xy, x^2, y^2$) for in-plane fitting. β is estimated with the least square fitting using the following normal equation:

$$\beta = (V^T V)^{-1} V^T \varphi \tag{4}$$

β represents the best linear unbiased estimator, assuming noise is uncorrelated. Similarly to Eq. 2, the conductivity is proportional to the second order derivative of the polynomial coefficients, corresponding to a summation of the polynomial coefficients of the second order terms:

$$\sigma = \frac{2}{\mu_0 \omega} \sum_{i=5,6} \beta_i \tag{5}$$

This fitting method can be extended by incorporating a weighting factor, which enhances the computation by focusing the fitting process on regions with similar signal intensities in MR contrast images. These weights, which can be derived from available MR contrast images, were calculated by comparing image magnitudes using a Gaussian distribution. This approach allowed for the computation of β coefficients through the weighted least squares solution. Using this method, the β coefficients were computed with the weighted least squares solution as:

$$\beta = (V^T W V)^{-1} V^T W \varphi$$
$$\text{where } W(r) = \frac{1}{\tau\sqrt{2\pi}} e^{-(|I(r)-I(r_0)|/2(\tau)^2)} \tag{6}$$

Here, τ represents the STD of the Gaussian distribution, while $I$ corresponds to the patched MR contrast images based on reconstruction kernel size used for the fitting operation. In this experiment, we set τ to 0.5. Considering the different regions of activation phase changes across different slices stimulated by different tasks, the kernel size was set to 17×17 for the finger-tapping task and 11×11 for the visual stimulation.

In the in-vivo experiments 2 and 3, to exclude the influence of regions outside of activation resulting from $B_1$ phase data, the ROI-based reconstruction was applied to the $B_1$ phase within the intersection area between the ROIs of fMRI and $B_1$ phase-based activations. Based on this ROI, Eq. (5) and (6) were computed with the kernel size=5×5.

For the comparison purpose in the simulation experiment 2, we additionally employed the phase-based cr-EPT[32] and integral-based EPT[33] methods. The integral-based EPT can be reformulated from Eq. (2) into an integral form, expressed as:

$$\sigma = (A2\mu_0\omega)^{-1} \cdot \int_C \nabla\varphi \, dl \tag{7}$$

where $A$ denotes the area enclosed by the curve $C$, which corresponds to the boundary defining the segmented ROI, such as CSF, WM, and GM. In this study, the kernel sizes were used as: 11×11, 15×15, 17×17, and 19×19.

On the other hand, the phase-based cr-EPT approach, which does not assume the piece-wise constant, explicitly includes the gradient of conductivity in its derivation, leading to a convection-reaction partial differential equation. The phase-based cr-EPT equation can be represented as:

$$-c\nabla^2\mathrm{p} + (\nabla\varphi \cdot \nabla\mathrm{p}) + \nabla^2\varphi\mathrm{p} - 2\omega\mu_0 = 0 \tag{8}$$

where $-c\nabla^2\mathrm{p}$ is the diffusion term, with $c$ being the constant diffusion coefficient, and p representing resistivity ($\mathrm{p} = 1/\sigma$). The convection term is given by $\nabla\varphi \cdot \nabla\mathrm{p}$, while $\nabla^2\varphi\mathrm{p} - 2\omega\mu_0$ constitutes the reaction term. In this study, the diffusion coefficient was set to 0.015, and the reconstruction kernel size were considered as: 5×5, 7×7, 9×9, and 11×11.

## Electro-magnetic Simulation Setting

We employed the Bessel boundary method[29] to generate a cylinder phantom with an infinite z-direction and the FDTD electro-magnetic (EM) simulation program, Sim4Life[30,31] (Zurich Med Tech, Zurich, Switzerland), to generate heterogeneous brain in-silico (Duke) data.

For the cylinder phantom simulation, the Bessel boundary method was used to compute the $B_1^+$ ($|B_1^+|$ and $\varphi^+$) and $B_1^-$ ($|B_1^-|$ and $\varphi^-$) fields in quadrature and anti-quadrature polarization modes at 128 [MHz]. The assigned electrical conductivity values for the default model were as follows[36,37,38]: the inner cylinder, assumed to be gray matter (GM) ($\sigma_{GM}$), was set to 0.5879 [S/m]; the outer cylinder, assumed to be white matter (WM) ($\sigma_{WM}$), was set to 0.3422 [S/m]. To investigate the relationship between the $B_1$ phase and reconstructed conductivity-based activation, we modified the conductivity value of the inner cylinder region to represent the activation region. The following models were considered in the simulation experiment 1: Model 1: $\sigma_{GM} + 0.06$, Model 2: $\sigma_{GM} + 0.04$, Model 3: $\sigma_{GM} + 0.02$, Model 4: $\sigma_{GM} - 0.02$, Model 5: $\sigma_{GM} - 0.04$, and Model 6: $\sigma_{GM} - 0.06$ [S/m]. Each cylinder simulation model was discretized with a voxel size of x = 2 [mm] and y = 2 [mm]. For the MR magnitude image synthesis, the bSSFP magnitude contrast was emulated to resemble the actual bSSFP image. Additionally, the transceive phase was obtained as the sum of $B_1^+$ and $B_1^-$ phase information (i.e., $\varphi^{\pm} = \varphi^{+} + \varphi^{-}$). The finalized image parameters were set as follows: a field of view (FOV) = 128×128 $\mathrm{mm}^2$, an image resolution = 2.0×2.0 $\mathrm{mm}^2$.

For the brain phantom simulation, the birdcage coil was operated in both quadrature and anti-quadrature polarization at 128 [Mhz]. The Duke human model from the IT'IS (Information Technologies in Society) Foundation was utilized in this study. For the FDTD computations, we assigned electrical conductivity values for the default model as follows[36,37,38]: CSF region ($\sigma_{CSF}$) = 2.1430 [S/m]; WM region ($\sigma_{WM}$) = 0.3422 [S/m]; GM region ($\sigma_{GM}$) = 0.5879 [S/m]. To investigate the effects of $B_1$ phase and conductivity changes versus activation in the heterogeneous brain phantom, we modified the conductivity value of an assigned region to mimic the activation state—the motor cortex, which was identified based on the gray matter region of the Duke phantom. The following models were designed for the simulation experiment 2: Model 7: $\sigma_{GM} - 0.02$, Model 8: $\sigma_{GM} - 0.04$, Model 9: $\sigma_{GM} - 0.06$, Model 10: region 1 = $\sigma_{GM} - 0.04$ and region 2 = $\sigma_{GM} + 0.04$, Model 11: region 1 = $\sigma_{GM} - 0.04$ and region 2 = $\sigma_{GM} + 0.02$, and Model 12: region 1 = $\sigma_{GM} - 0.04$ and region 2 = $\sigma_{GM} + 0.01$ [S/m]. Based on these scenarios, the FDTD simulation adopted a discretized voxel size of x = 2 [mm], y = 2 [mm], and z = 2 [mm]. For the MR magnitude synthesis, the bSSFP magnitude contrast was emulated to resemble the actual bSSFP image. Additionally, the transceive phase was obtained as the sum of $B_1^+$ and $B_1^-$ phase information (i.e., $\varphi^{\pm} = \varphi^{+} + \varphi^{-}$). The finalized image parameters were set as follows: a field of view (FOV) = 128×128 [$\mathrm{mm}^2$], an image resolution = 2.0×2.0 [$\mathrm{mm}^2$], a slice thickness = 2 [mm].

Based on the synthesized bSSFP data for both the cylinder and brain models, two block design simulations were formulated. In block design 1 (indicated by the black colored box in Fig. 4B and Fig. 5A), the simulated default model was utilized, while in block design 2 (indicated by the white colored box in Fig. 4B and Fig. 5A), the simulated activation model was employed. The structure of these simulation block designs mirrored that of the in-vivo experiment, comprising a total of 80 dynamics. To capture the simulated brain contrast, a synthetic MPRAGE image was additionally generated using the segmentation information from the cylinder and brain models.

# Supplementary Materials

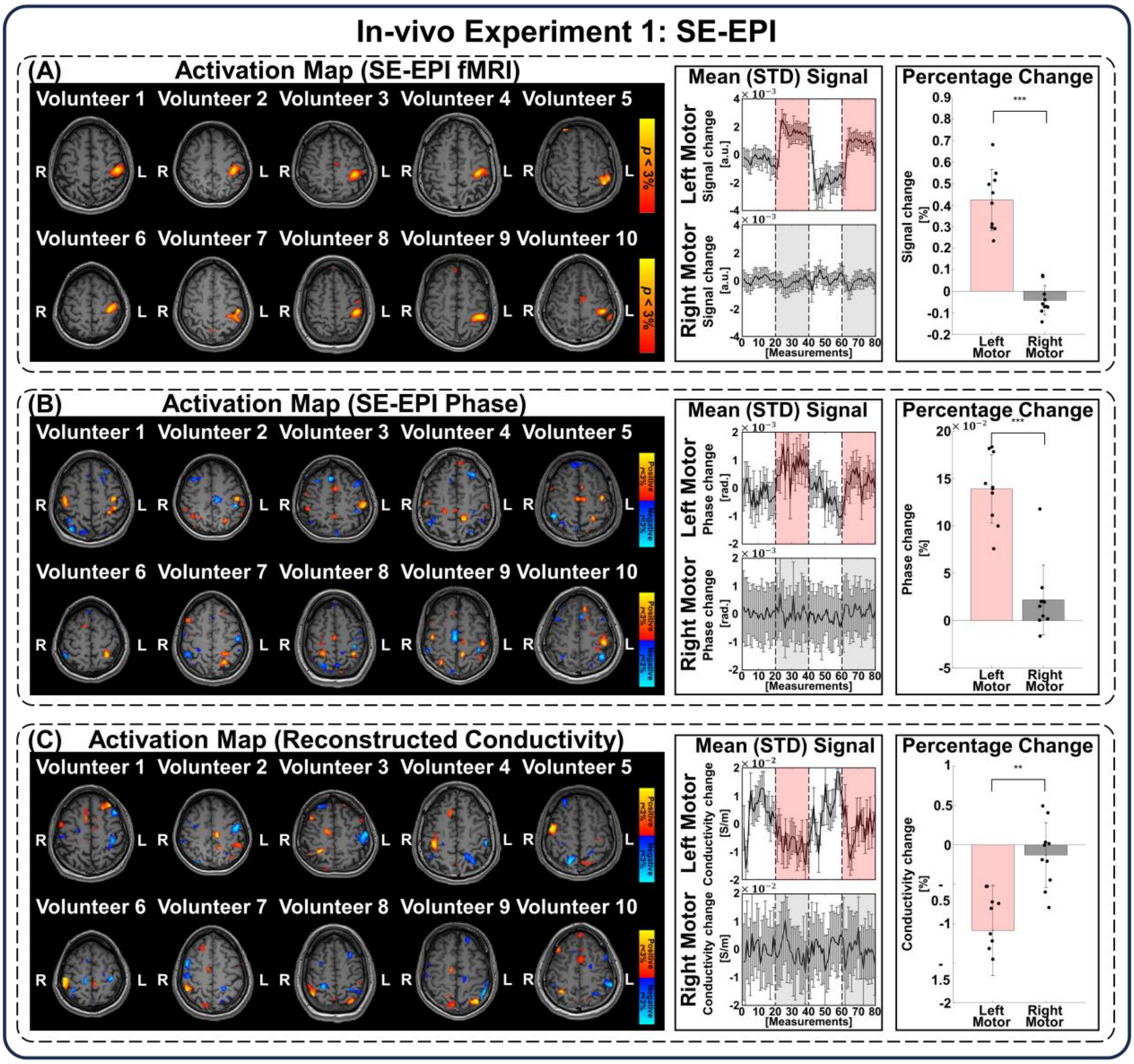

**Fig. S1|** In-vivo experiment 1: Functional MRI, phase, and reconstructed conductivity activation maps, acquired using SE-EPI, with mean (STD) signal and activation percentage during finger-tapping stimulus in target regions ($n$=10). For the percentage change analysis: in the fMRI, statistical significance is indicated by asterisks: *** for $p$<0.001, ** for $p$<0.01; in the functional activity changes of phase and conductivity, by asterisks: *** for $r$<0.001, ** for $r$<0.01. **A** SE-EPI fMRI activation map. The activation regions were identified at a significance level of $p$<0.03. Mean (STD) temporal-series signal derived based on right (control) and left motor (activation) regions. **B** Phase activation map obtained from SE-EPI. **C** Reconstructed conductivity activation map obtained from SE-EPI phase. For the phase-based conductivity reconstruction, the weighted polynomial fitting method was utilized.

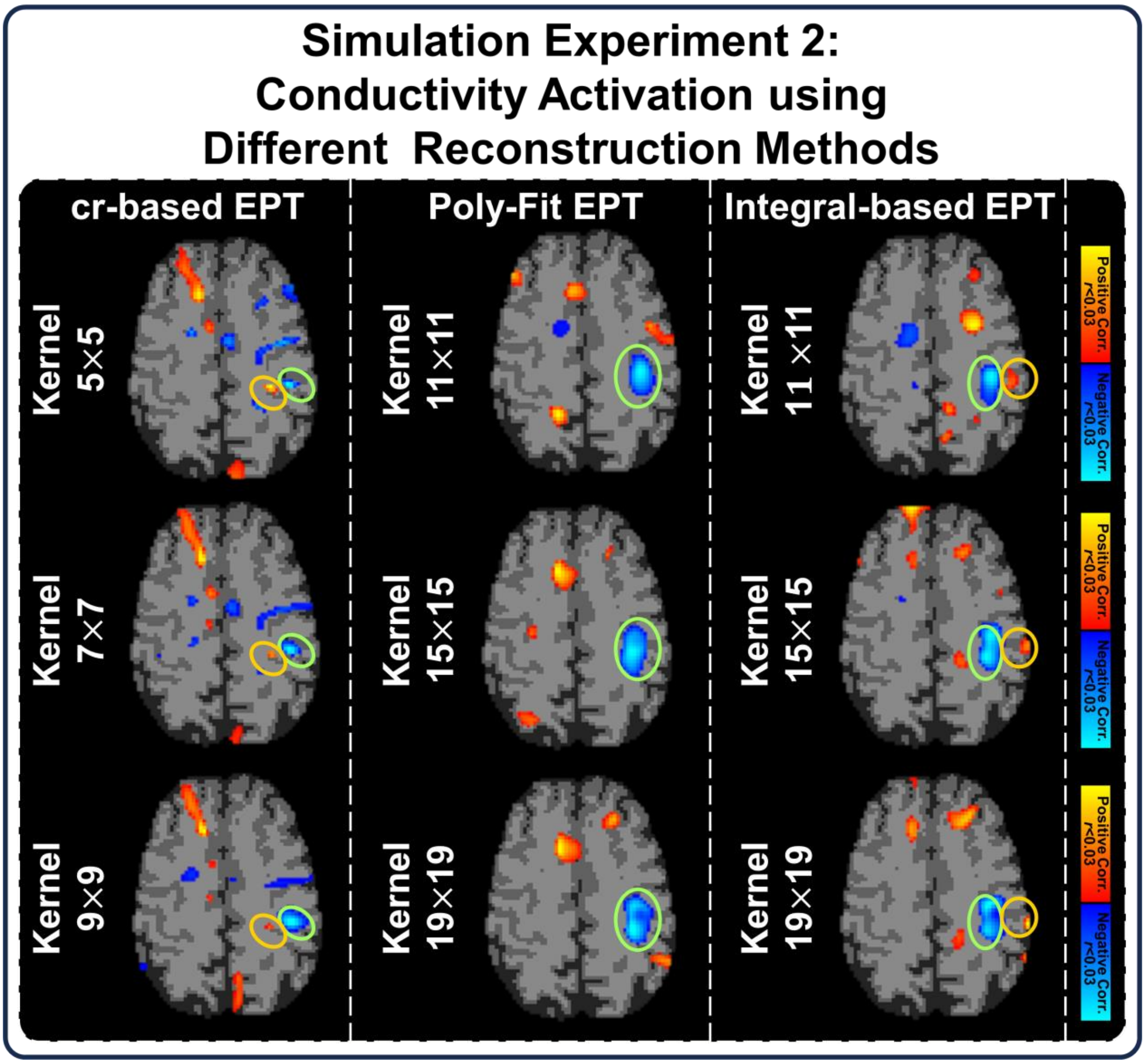

**Fig. S2|** Phase-based conductivity reconstruction results from simulation experiment 2 (Model 8) using different reconstruction methods and kernel sizes. In cr-based EPT reconstructions, the use of a small kernel led to spurious positive correlations in conductivity activation within the activation region (orange circle). In contrast, the Poly-Fit EPT method did not produce spurious conductivity activation around the activation region (green circle) even with smaller kernels. However, as the kernel size decreased, spurious signals appeared in various other regions. The integral-based method generated spurious activation signals around the activation region regardless of the kernel size (orange circle).

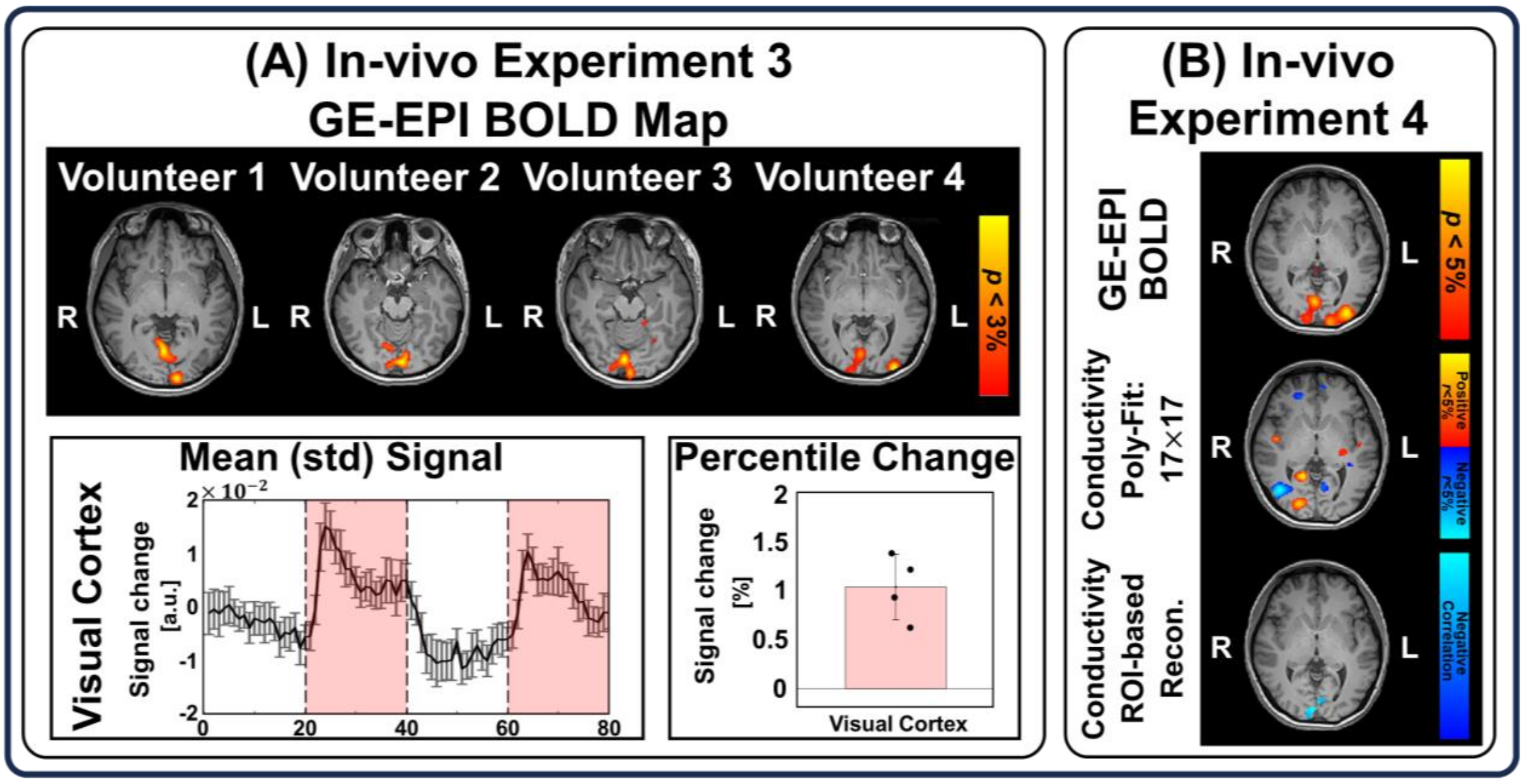

**Fig. S3|** Additional results for in-vivo experiments 3 and 4. **A** In-vivo experiment 3: BOLD GE-EPI maps with mean (STD) signal and activation percentage during visual stimulus (*n*=4). The BOLD signals were detected at a significance level of $p<0.03$. Mean (STD) temporal-series signal derived based on the visual cortex regions. **B** In-vivo experiment 4: Activation maps from GE-EPI BOLD ($p<0.05$) (top) and conductivity ($r<0.05$); Poly-Fit result with a kernel size of 17×17 (middle), and ROI-based reconstruction (bottom).

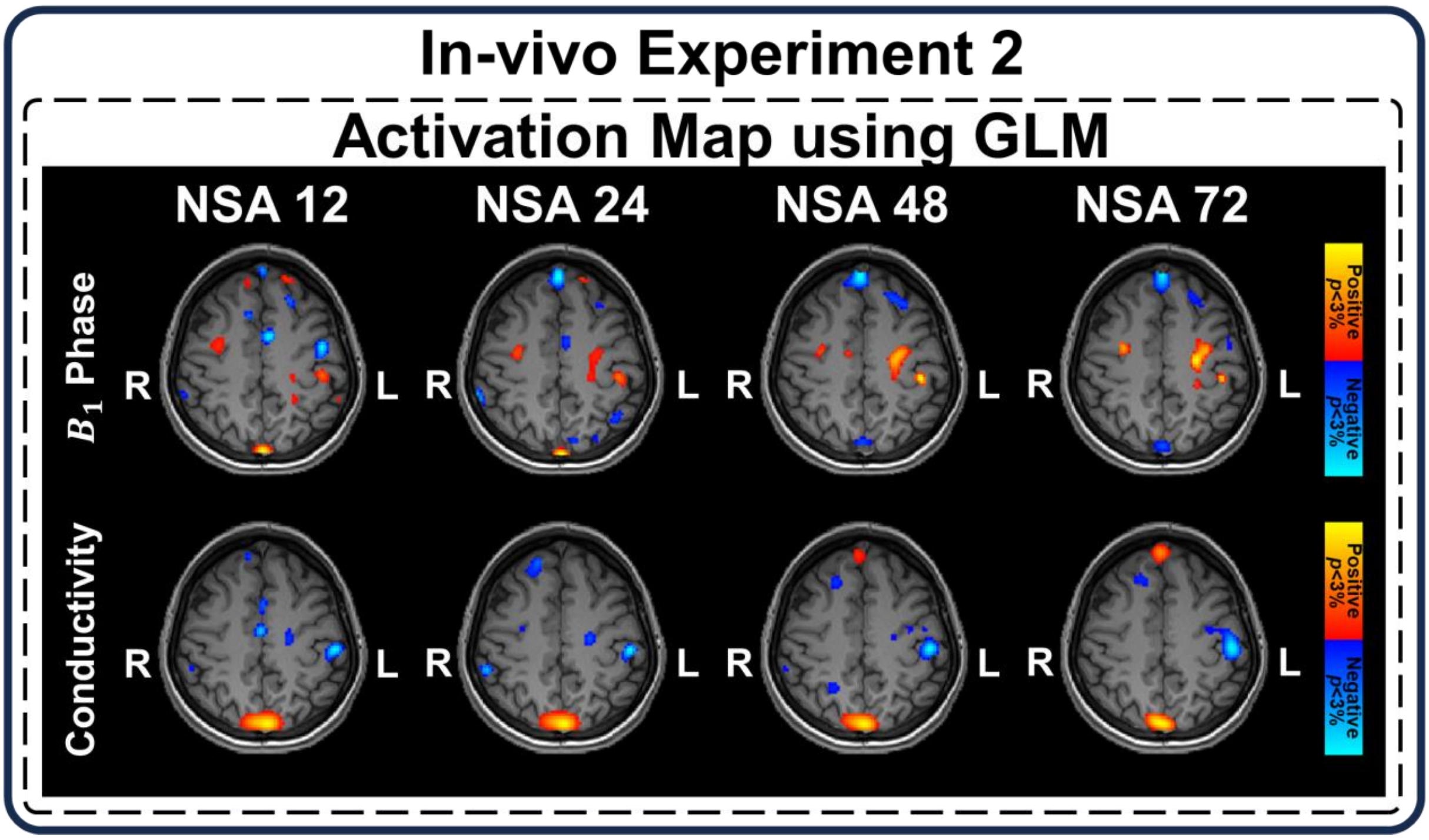

**Fig. S4|** Additional activation map using GLM. The results of activation map computation using GLM ($p$<0.03) for in-vivo experiment 2 (NSA = 12, 24, 48, and 72). The results with GLM exhibit a trend similar to the correlation method (see Fig. 3); however, for NSA values below 24, the phase activation showed a weaker activation amplitude near the left motor cortex. For NSA values above 48, no significant differences in activation region detection were observed when compared to the correlation method.

# Data availability

Simulation: due to the use of the commercial FDTD software (Sim4Life) with the commercial head models (Duke) to compute electromagnetic fields in this study, the simulation dataset cannot be shared.

In-vivo: the pre-processed data (central recentering phase correction and registration) can be provided upon email request to the corresponding author after the submitted journal's publication.

# Code availability

The SPM12, FSL, Python, and Matlab (MATLAB, MathWorks, USA) were used to process the dataset. The related image processing codes can be provided upon direct email request to the corresponding author after the submitted journal's publication.